\documentclass[a4paper,12pt]{article}
\usepackage{epstopdf}
\usepackage{color}
\usepackage{cite}
\usepackage{amsmath,amssymb}
\input{colordvi.tex}
\usepackage{graphicx}
\usepackage{comment}
\usepackage{soul} % to enable \sout for strike-through text

\begin{document}

\newcommand{\lsim}{\stackrel{<}{_\sim}}
\newcommand{\gsim}{\stackrel{>}{_\sim}}

\newcommand{\Ms}{\mathrm{M}_\odot}

\renewcommand{\theequation}{\thesection.\arabic{equation}}
\renewcommand{\thefootnote}{\fnsymbol{footnote}}
\setcounter{footnote}{0}
\def\thefootnote{\fnsymbol{footnote}}

\begin{titlepage}

\begin{center}

\vskip .75in

{\Large \bf Structure Formation with Two Periods of Inflation: \vspace{2mm} \\ Beyond PLaIn $\Lambda$CDM} \vspace{10mm} \\

\vskip .75in

{\large
Kari~Enqvist$\,^{a}$, Till Sawala$\,^{a}$, and Tomo~Takahashi$\,^b$
}

\vskip 0.25in

{\em
$^{a}$
Department of Physics and
Helsinki Institute of Physics, FIN-00014, \\
 University of Helsinki, Finland
\\
$^{b}$
Department of Physics, Saga University, Saga 840-8502, Japan
}

\end{center}
\vskip .5in

\begin{abstract}

We discuss structure formation in models with a spectator field
  in small-field inflation which accommodate a secondary period
  of inflation.  In such models, subgalactic scale primordial
fluctuations can be much suppressed in comparison to the usual
power-law $\Lambda$CDM model while the large scale fluctuations remain
consistent with current observations. We discuss how a secondary
inflationary epoch may give rise to observable features in the small
scale power spectrum and hence be tested by the structures in the
Local Universe.

\end{abstract}

\end{titlepage}

\renewcommand{\thepage}{\arabic{page}}
\setcounter{page}{1}
\renewcommand{\thefootnote}{\#\arabic{footnote}}
\setcounter{footnote}{0}

%%%%%%%%%%%%%%%%%%%%%%%%%%%%%%%%%%%%%%%%%%%%%%%%%%
\section{Introduction} \label{sec:intro}
%%%%%%%%%%%%%%%%%%%%%%%%%%%%%%%%%%%%%%%%%%%%%%%%%%
The standard ``Lambda Cold Dark Matter'' (hereafter {\it plain}
$\Lambda$CDM, or $\Lambda$CDM with {\it POwer LAw Inflation}\footnote{
Here ``power law" indicates that  primordial power spectrum is of  a ``power-law" form. 
Do not confuse with power law inflation  model in which the inflaton potential is given by an exponential function.
}) model of
cosmology is founded on three key assumptions: firstly, that the
Universe contains, in addition to baryonic matter, a collisionless,
``cold'' dark matter component that accounts for $\sim 5/6$ of the
matter density; secondly, that the expansion of the Universe is
described by the Friedmann equations with a cosmological constant
term, $\Lambda$, and, thirdly, the implicit assumption that structure
formation has its origin in the primordial perturbations seeded by
inflation.

The plain $\Lambda$CDM model has successfully predicted structure
formation over many epochs and orders of magnitude, from the
structures observed in the Cosmic Microwave Background (CMB) at
$z=1100$ to the sizes and distributions of clusters and galaxies at
much later time. However, tensions between the CMB and measurements on
small scales have also been reported. These have either been
attributed to poorly understood systematic effects in various data
sets, or interpreted as an indication of new physics.

On small scales, predictions for structure formation concern the
abundance and internal structure of low mass dark matter halos in the
Local Universe. Here, observations of the Milky Way and Andromeda
satellite populations in particular, appear to be in disagreement with
N-body simulations based on the plain $\Lambda$CDM model.

This tension is twofold. The ``missing satellites'' problem
\cite{Klypin:1999uc, Moore:1999nt} refers to the apparent paucity of
luminous satellite galaxies compared to the large number of dark
matter substructures predicted in plain $\Lambda$CDM. While processes
including supernova feedback (e.g. \cite{Larson:1974}), and cosmic
reionisation (e.g. \cite{Efstathiou:1992}) are expected to reduce the
number of observed satellite galaxies, the apparent excess of
substructures in the plain $\Lambda$CDM is not limited to the lowest
mass scales: simulations also predict the presence of subhalos so
massive that they should not be affected by reionisation (and hence
deemed ``too big to fail'', \cite{BoylanKolchin:2011de}), but whose
internal structure seems incompatible with that of the brightest
observed satellites.

Recent hydrodynamic simulations have shown that, when baryonic effects
are included, the observed Local Group satellite populations can be
reproduced in $\Lambda$CDM (e.g. \cite{Sawala:2015cdf}), but only if
the Local Group mass is $\sim 2-3\times 10^{12}\Ms$, at the lower end
of the observational limits \cite{Gonzalez:2013pqa}. It has also been
shown that, if the CDM assumption of the plain $\Lambda$CDM model is
relaxed, simulations of warm dark matter \cite{Lovell:2016fec,
  Bozek:2018ekc} and photon-coupled dark matter
\cite{Schewtschenko:2015rno}, which both reduce the small-scale power
after decoupling, give a better agreement to satellite kinematics than
the equivalent plain $\Lambda$CDM models. Conversely, observed
structures in the Lyman-$\alpha$ forest \cite{Viel:2013apy,
  Irsic:2017ixq}, and the abundance of Milky Way satellites
\cite{Kennedy:2013uta} also provide upper limits for any reduction of
small scale power relative to plain $\Lambda$CDM.

Warm dark matter models prevent the hierarchical formation of
structures below the free-streaming scale (e.g. \cite{Bode:2000gq}),
but above that, the abundance of structures relative to CDM differs
significantly over only a narrow range. Consequently, within current
limits, WDM effects are testable only at or below the scale of
ultra-faint dwarf galaxies \cite{Lovell:2016fec}. This gives WDM
models limited predictive power. Furthermore, the motivation for the
required WDM particle appears somewhat ad-hoc. Another, perhaps more
attractive possibility, is to modify the power spectrum of primordial
perturbations. The modification should be such that it persists over
$\sim 4$ orders of magnitude; hence e.g. local features in the
inflation potential are unlikely to do the job.

Instead, here we propose a unified description of structure formation
that relies on two periods of inflation and accounts both for the
power on CMB scales and for the apparent suppression of power on
subgalactic scales.

There exist models for two periods of inflation making use e.g. of a
suitably arranged inflaton potential with an intervening phase
transition \cite{Kamionkowski:1999vp,Yokoyama:2000tz}\footnote{
It has been argued that the CMB spectral $\mu$ distortion would be
useful to differentiate models with a small-scale suppression of the
matter power spectrum due late-time effects (such as different dark
matter properties) from those caused by a modification of the
primordial power spectrum \cite{Nakama:2017ohe}.
}.

In this paper we consider a model where the inflaton field, giving
rise to a slow roll inflation, is complemented by another scalar
field, which is dynamically irrelevant during the first period of
inflation. Such a scalar is called a spectator field. We know that
spectator fields exist: the Higgs field is one example (see, e.g.,
\cite{Enqvist:2014bua,Enqvist:2014tta,Enqvist:2015sua} for
fluctuations of the mean Higgs field during inflation), barring the
possibility that the Higgs field itself is the inflaton. Another, much
studied spectator field is the curvaton
\cite{Enqvist:2001zp,Lyth:2001nq,Moroi:2001ct} which imprints the
perturbations it receives during inflation on radiation and matter
after inflation\footnote{Another example would be the modulated
  reheating model \cite{Dvali:2003em,Kofman:2003nx}}.  Rather than the
inflaton, the curvaton could then be the origin of the whole of the
observed spectrum of perturbations.

Here we do not assume that the spectator field contributes to the
primordial perturbation in a significant manner. Instead, we assume
that it gives rise to second period of inflation, which happens under
the conditions to be discussed below. Such a second period of
inflation affects the power spectrum both on the CMB scales as well as
on subgalactic scales. In this paper we shall discuss a model with
two periods of inflation that yields the desired suppression at small
scales while remaining in agreement with the observed properties of
the CMB power spectrum, making it possible to address the apparent
shortfalls of the plain $\Lambda$CDM model at small scales in
completely different, and astrophysically decoupled regimes. Such
agreement, however, makes certain demands on the inflaton model; not
all inflationary potentials are consistent with two inflationary
periods. We will also demonstrate that for our working example, the
fact that some subgalactic structure has actually been observed gives
rise to a lower limit on the number of $e$-folds of the first
inflationary period.

This paper is organized as follows. In Section~\ref{sec:spectator}, we
briefly introduce the general concept of a spectator field, and in
Sections~\ref{sec:background} and \ref{sec:perturbation} we describe
the background evolution and perturbation, respectively. In
Section~\ref{sec:model}, we introduce a concrete inflation model. In
Section~\ref{sec:small} we discuss the predictions for small scale
power. We conclude with a summary in Section~\ref{sec:summary}.

%%%%%%%%%%%%%%%%%%%%%%%%%%%%%%%%%%%%%%%%
\section{Spectator fields}\label{sec:spectator}
%%%%%%%%%%%%%%%%%%%%%%%%%%%%%%%%%%%%%%%%%%

Fields whose energy densities during inflation are irrelevant for the
expansion rate of the Universe are called spectators. If they are
light, both their mean field values and the field perturbations will
be subject to inflaton-driven inflationary expansion.  At the onset of
inflation the spectator energy density is subdominant. The spectator
field $\sigma$ is assumed to begin to oscillate during the
radiation-dominated period after the inflaton field $\phi$ has decayed
(or during the period dominated by the oscillations of the
inflaton). Its energy density depends on the initial spectator field
value $\sigma_\ast$, as well as on the exact form of the spectator
potential. During inflation, the mean spectator field is subject to
fluctuations and in a given inflationary patch is one realization of
the probability distribution, which is determined by the Fokker-Planck
equation\footnote{Assuming slow-roll inflation, as will be done here.}
as was first pointed out by Starobinsky \cite{Starobinsky:1986fx} (for
a discussion in the context of spectators, see e.g
\cite{Enqvist:2012xn,Hardwick:2017fjo}).

At the end of inflation, the spectator square-mean-field has a value
$\sigma_\ast$, which serves as the initial spectator field value for
its subsequent evolution. As a consequence of its stochastic evolution
during inflation, it may have a value greater than the Planck mass
$M_{\rm pl}$. If inflation lasts long enough, the probability
distribution for the initial curvaton field equilibrates; otherwise
there will be a dependence on the curvaton field value before the
onset of inflation (however, equilibration is not automatic; for a
recent discussion on the fluctuations of spectator fields, see
\cite{Hardwick:2017fjo}). In any case, if $\sigma_\ast > M_{\rm pl}$,
the spectator may end up dominating the Universe even before it starts
to oscillate. This happens, provided that the spectator is still
slowly rolling down its potential, whence a period of secondary
inflation can
ensue\cite{Langlois:2004nn,Moroi:2005kz,Ichikawa:2008iq,Dimopoulos:2011gb}. Thus,
very crudely, first there is a period of inflation driven by the
``usual" inflaton field; then the inflaton decays; after a while, the
energy density of the slowly rolling spectator becomes dominant and
generates a second period of inflation, which ends when the spectator
decays. This scenario, which we investigate in the present paper, has
some very interesting consequences, in particular for the spectrum of
perturbations at small scales. As we will discuss, these consequences
will also depend on the details of the slow roll inflation model.

Depending on the duration of the secondary inflation driven by the
spectator, the current observable scales (such as CMB) may have exited
the horizon either during the primary slow roll inflationary phase or
during the secondary, spectator-driven phase.  In the standard case
with no inflating spectator, the required number of $e$-folds is
usually taken to be $N \sim 50 - 60$ $e$-folds.  However, depending on
the duration of the secondary inflationary phase, we may tolerate
primary $e$-folds as low as $N \sim 10$. In such a case, the
predictions for the primordial power spectrum can be drastically
modified.

The spectator modifies the spectrum of perturbations already at the
CMB scales. Because the first phase of inflation ends early, the modes
corresponding to the CMB scales which exited the horizon closer to the
end of the inflaton period, at which the inflaton starts to move
faster, even if it is still slowly rolling. As a result, there will be
a large running of the spectral index, which may also significantly modify
the predictions for astrophysical scales.

%%%%%%%%%%%%%%%%%%%%%%%%%%%%%%%%%%
\subsection{Background evolution}\label{sec:background}

Before describing primordial density fluctuations, let us first look
at the background evolution when a primary inflationary period is
followed by a secondary period driven by a spectator.

If the spectator is to drive a secondary phase of inflation, it has to
dominate the Universe before it begins to oscillate.  This condition
is given by
\begin{equation}
%\label{ }
U(\sigma_\ast)  \ge \rho_r (t_{\rm osc}),
\end{equation}
where $\sigma_\ast$ is the spectator field value set during the first
inflationary phase, $\rho_r (t_{\rm osc})$ is the radiation energy
density at the beginning of the spectator oscillation $t=t_{\rm osc}$
and $U(\sigma)$ is the potential of $\sigma$.  The spectator
oscillation starts when $H \sim m_\sigma$, where $m_\sigma$ is the
mass of the spectator. Throughout this paper we will assume that the
spectator potential reads simply
\begin{equation}
\label{curvpotential}
U(\sigma) = \frac 12 m_\sigma^2 \sigma^2~,
\end{equation}
whence the above condition can be
rewritten as
\begin{equation}
\label{condforinf}
\frac12 m_\sigma^2 \sigma_\ast^2 \ge 3 M_{\rm pl}^2 m_\sigma^2,
\end{equation}
where $ M_{\rm pl}$ is the reduced Planck mass. From
(\ref{condforinf}) we obtain the condition for the spectator-driven
secondary inflation as
\begin{equation}
%\label{ }
\sigma_\ast \gtrsim \sqrt{6} M_{\rm pl}.
\end{equation}

The mass and the initial field value of the spectator are generally
assumed to be model parameters.  If one has a long period of initial
inflaton-driven inflation so that the curvaton reaches the
Fokker-Planck equilibrium distribution, a typical value of the
amplitude of the spectator field is given by
\cite{Enqvist:2012xn,Hardwick:2017fjo}
\begin{equation}
\label{eq:equil_value}
\sigma_{\ast} \simeq \frac{H_\ast^2}{m_\sigma},
\end{equation}
with $H_\ast$ being the Hubble rate during inflation, in which
de-Sitter (a constant $H_\ast$) background is assumed.  Although
whether this distribution can be reached or not depends on inflation
models, when an inflation model with a potential of the plateau type
is adopted, the spectator field can obtain the de-Sitter equilibrium
\cite{Hardwick:2017fjo}.  On the other hand, for the case of
  large-field inflation models, the equilibrium solution would not be
  reached although the spectator can still acquire a super-Planckian
  amplitude $\sigma_\ast > M_{\rm pl}$ in the regime of an eternal
  inflation \cite{Hardwick:2017fjo}. 

Once the equilibrium value is reached, by using the fact that the
inflationary Hubble scale is directly related to the slow-roll
parameter $\epsilon$ as $H_\ast^2 / M_{\rm pl}^2 \simeq 1.6 \times
10^{-7} \epsilon$\footnote{
Here we assume that primordial perturbations are generated only from
an inflaton.
},
the above expression can be recast as
\begin{equation}
\label{eq:sigma_ast}
\sigma_\ast \sim 8 M_{\rm pl} \left( \frac{10^{6}~{\rm GeV}}{m_\sigma} \right)  \left( \frac{\epsilon}{2 \times 10^{-5}} \right) \,,
\end{equation}
from which one can see that the amplitude of a spectator field can be
as large as $\sigma_\ast > {\cal O}(M_{\rm pl})$ for some set of parameters.  As
noted before, when $\sigma_\ast > M_{\rm pl}$, the spectator can give
rise to the secondary inflation. From Eq.~\eqref{eq:sigma_ast} we can
see that such a scenario may in fact be generic, given the stochastic
behaviour of $\sigma$ during the primary inflation and the relation
\eqref{eq:equil_value}.
Even if the equilibrium value is not reached,  the initial value of the spectator can be regarded as 
a model parameter,  which does not prohibit $\sigma_\ast$ of having a value as large as $\sigma_\ast > {\cal O}(M_{\rm pl})$.  
Therefore a scenario with a second period of inflation driven by a spectator could be realized in a broad class of model.

\begin{figure}[htbp]
\begin{center}
\vspace{10mm}
\resizebox{120mm}{!}{
     \includegraphics{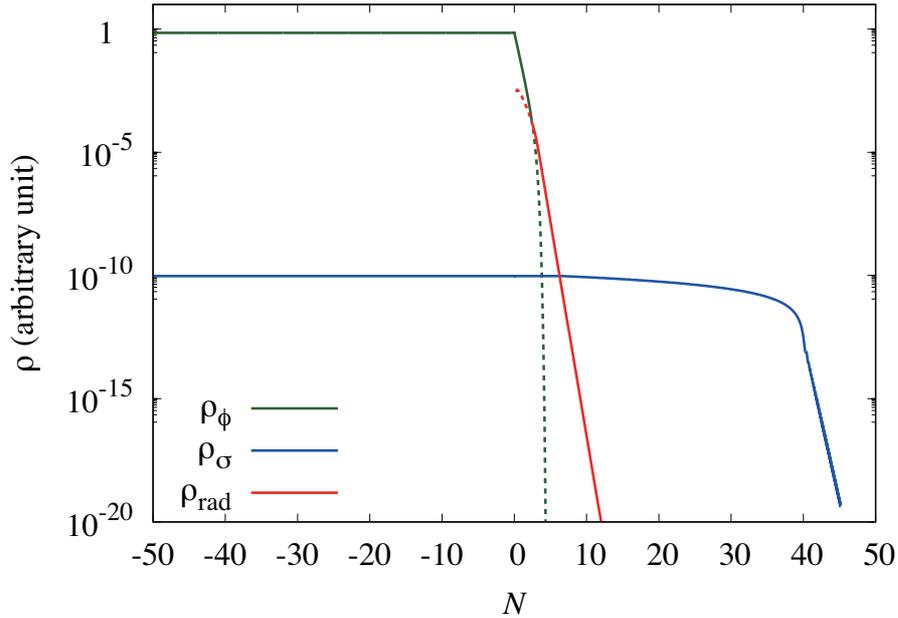}
}
\end{center}
\caption{Evolution of energy densities of the inflaton field,
  spectator field and radiation in models where the secondary
  inflation is driven by a spectator.  The number of $e$-folds $N$ is
  normalized to 0 at the end of the first inflation in this figure.
  Depending on the decay rate of the inflaton, the Universe may have
  been dominated by oscillating $\phi$ from the end of the first
  inflation to the beginning of the second inflation. In this case,
  there is no intermediate radiation-dominated phase.}
\label{fig:rho_evolution}
\end{figure}

In Fig.~\ref{fig:rho_evolution}, we show a typical thermal history in
the scenario with two periods of inflation, the second driven by a
spectator. We display the evolution of energy densities of the
inflaton, the spectator and radiation, denoted as $\rho_\phi,
\rho_\sigma$ and $\rho_{\rm rad}$, respectively.  In the following, we
focus on the case where the spectator gives a second inflationary
period after the CMB scales have already exited the horizon close to
the end of the first period of inflation so that the small scale
fluctuations are suppressed.

%%%%%%%FIGURE%%%%%%%%%%%%%%%%%%%%%%
%%%%%%%%%%%%%%%%%%%%%%%%%%%%%%%%%

%%%%%%%%%%%%%%%%%%%%%%%%%%%%%%%%%%%%%%%%%%%%%%%%%%
\subsection{Density perturbation}\label{sec:perturbation}

In a scenario where a spectator field is present, both the inflaton
and the spectator may contribute to the primordial fluctuations (in
which case the spectator is a curvaton) for the mode which exited the
horizon during the first inflationary epoch driven by the inflaton.
For the modes which exited the horizon during the second inflationary
epoch driven by a spectator, primordial fluctuations are sourced from
the spectator alone.  Although we consider a case where fluctuations
from a spectator field can be neglected in the next section, here we
also discuss the curvature perturbation generated by a spectator in
order to clarify in what case spectator fluctuations can be
neglected. Here we consider the curvaton type model.

As mentioned above, for modes which exited the horizon during the
first inflationary period generated by the inflaton field, the
curvature perturbation $\zeta$ is in general given by the sum of two
contributions\footnote{
For a general discussion on a scenario where the inflaton and the
spectator can both contribute to the curvature perturbation, we refer
the readers to
\cite{Langlois:2004nn,Lazarides:2004we,Moroi:2005kz,Moroi:2005np,Ichikawa:2008iq,Fonseca:2012cj,Enqvist:2013paa,Vennin:2015vfa,Fujita:2014iaa,Haba:2017fbi}
for the curvaton model and \cite{Ichikawa:2008ne} for modulated
reheating scenario.
}: 
\begin{equation}
%\label{ }
\zeta_I  = \zeta^{(\phi)}_I + \zeta^{(\sigma)}_I,
\end{equation}
where $\zeta^{(\phi)}_I$ and $\zeta^{(\sigma)}_I$ are respectively the
curvature perturbations generated from the inflaton and the spectator
(which in this case should be called the curvaton).  A subscript $I$
indicates that the perturbations correspond to modes which exited the
horizon during the first inflation.  The inflaton part can be written
as
\begin{equation}
\label{eq:zeta_phi1}
\zeta^{(\phi)}_I =  \frac{1}{M_{\rm pl}^2} \frac{V (\phi)}{V_\phi (\phi) } \delta \phi_\ast
= \frac{1}{\sqrt{ 2 \epsilon M_{\rm pl}^2}} \delta \phi_\ast,
\end{equation}
in which $V(\phi)$ and $V_\phi (\phi)$ are the potential for the
inflaton and its derivative with respect to the inflaton field $\phi$
and the fluctuation of the inflaton is given by $\delta \phi_\ast =
H_\ast /(2\pi)$.  Therefore the primordial power spectrum sourced by
inflaton fluctuations is given by
\begin{equation}
\label{eq:P_prim_inflaton}
{\cal P}_{\zeta}^{(\phi)} (k) = \frac{1}{12\pi^2 M_{\rm pl}^6} \frac{V(\phi)^3}{V_\phi(\phi)^2} \,.
\end{equation}

For the curvaton part, here we give the expression for the case where
the curvaton drives the secondary inflation since we mainly consider
this kind of scenario in this paper.  By adopting the $\delta N$
formalism \cite{Starobinsky:1986fxa,Sasaki:1995aw,Sasaki:1998ug}, the
curvature perturbation generated from the spectator is given by
$\zeta^{(\sigma)} = (\partial N /\partial \sigma_\ast) \delta
\sigma_\ast$ where $N$ is the number of $e$-folds from the epoch when
the mode exited the horizon to that at the decay of the curvaton.
However, in the case where the second inflation is driven by a
spectator, the $\sigma_\ast$-dependence of $N$ comes from the period
during the second inflation. The number of $e$-folds during such a
secondary inflation is given by \cite{Ichikawa:2008iq}
\begin{equation}
%\label{ }
N_2 = - \frac{1}{M_{\rm pl}^2} \int_{\sigma_\ast}^{\sigma_{\rm end}} \frac{U(\sigma)}{U_\sigma (\sigma)} d\sigma,
\end{equation}
where $U(\sigma)$ and $U_\sigma (\sigma)$ are the potential of the
curvaton and the derivative with respect to $\sigma$, respectively.
Here $\sigma_{\rm end}$ is the value of $\sigma$ at the end of the
second inflation.  Here we adopt the potential \eqref{curvpotential}
and hence $N_2$ is given by
\begin{equation}
%\label{ }
N_2 =  \frac{1}{4 M_{\rm pl}^2} (\sigma_\ast^2 - \sigma^2_{\rm end} ),
\end{equation}
from which we can obtain \cite{Ichikawa:2008iq}
\begin{equation}
\label{eq:zeta_inf_cur}
\zeta^{(\sigma)}_I  = \frac{\sigma_\ast}{2M_{\rm pl}^2} \delta \sigma_\ast,
\end{equation}
with $\delta \sigma_\ast = H_\ast /(2\pi)$.  The condition where
fluctuations from the curvaton are negligible can be written as
\begin{equation}
%\label{ }
\frac{\zeta^{(\sigma)}_I}{\zeta^{(\phi)}_I} = \sqrt{2 \epsilon} \frac{\sigma_\ast}{M_{\rm pl}} \ll 1, 
\end{equation}
from which one can see that, even when $\sigma_\ast > M_{\rm pl}$,
fluctuations from a spectator can be negligible if small-field
inflation models with very small $\epsilon$ are assumed.

On small scales where modes exited the horizon during the second
inflation driven by a spectator, the curvature perturbation is given
by the same expression as for the ones generated from the inflaton
during the first inflationary period. Hence one can write
\begin{equation}
%\label{ }
\zeta^{(\sigma)}_{II} = \frac{1}{M_{\rm pl}^2} \frac{U}{U_\sigma} \delta \sigma_{\ast II}
\simeq 3 \frac{m_\sigma}{M_{\rm pl}} \left( \frac{\sigma_\ast}{10 M_{\rm pl}} \right)^2,
\end{equation}
with $\delta \sigma_{\ast II} = H_{\ast II} / 2 \pi$ being determined
by the Hubble parameter during the second inflation~$H_{\ast II}$.
The mass for the spectator should be much smaller than $ H_\ast (< M_{\rm pl}) $
since otherwise the spectator field plays the role of the inflaton,
and hence $\zeta^{(\sigma)}_{II} \ll 10^{-5}$. Therefore primordial
fluctuations on small scales would be much smaller than those
generated on large scales from the inflaton fluctuations. This gives
an effective cutoff in the power spectrum at some scale, which may
have interesting implications for small scale structure while CMB
scale is not affected. In the next section, we assume the above kind
of scenario in an inflaton model with very small $\epsilon$.

%%%%%%%%%%%%%%%%%%%%%%%%%%%%%%%%%%%%%%%%%%%%%%%%%%
\section{A concrete model and power spectrum}\label{sec:model}

To compute the power spectrum in this kind of scenario explicitly, we
also need to specify the inflaton model. Although many inflationary
potentials would be admissible to have a model with a consistent $n_s$
and $r$ with current observations, here let us consider the following
inflaton model, which is a hybrid inflation with a fractional power
whose potential is given by
\begin{equation}
\label{eq:V_VHI}
V(\phi) = V_0 \left( 1+ \left( \frac{\phi}{\mu} \right)^p \right),
\end{equation}
where $V_0$ represents the scale of inflation, which is determined by
the normalization condition.  Here $\mu$ and $p$ are model parameters
which will be chosen to give the spectral index $n_s$ consistent with
current observations. This type of inflation model is also called
``valley hybrid inflation (VHI)" in \cite{Martin:2013tda}.

The slow-roll parameters in this model are given by 
\begin{equation}
\label{eq:slow-roll}
\epsilon = \frac{p^2}{2} \left( \frac{M_{\rm pl}}{\mu} \right)^2
\frac{ x^{2p-2}}{ (1+x^p)^2}, \qquad \eta = p (p-1) \left(
\frac{M_{\rm pl}}{\mu} \right)^2 \frac{ x^{p-2}}{ 1+x^p},
\end{equation}
where we have defined 
\begin{equation}
%\label{ }
x \equiv \frac{\phi}{\mu}.
\end{equation}
The number of $e$-folds counted from the end of the first inflationary
period is given by
\begin{equation}
%\label{ }
  N = \frac1p \left( \frac{\mu}{M_{\rm pl}} \right)^2
  \left[ \frac{1}{2-p} (x_\ast^{2-p} - x_{\rm end}^{2-p} ) + \frac12 (x_\ast^2 - x_{\rm end}^2 ) \right].
\end{equation}

Now we consider the case of $p <1$, which corresponds to $n_s < 1$,
however when we assume $N = 50 - 60$, $n_s$ is very close to unity.
As described below, if the secondary inflation is driven by a
spectator field, $N$ can be much reduced, which can lead to a value
consistent with current observations ($n_s \sim 0.96$).  To consider
the situation where the spectator does not contribute to primordial
fluctuations but only affects the background dynamics, we require a
very small value for $\epsilon$ as shown in the previous section. To
realize this, here we assume a large value for $\mu$ i.e. $\mu \gg
M_{\rm pl}$ (for concreteness, we assume $\mu = 10^3 M_{\rm pl}$ in
the following). We also need a very small $x = \phi/\mu \ll 1$.  In
this case, the approximate number of $e$-folds is
\begin{equation}
\label{eq:N_hybrid_fractional}
N = \frac{1}{p(2-p)} \left( \frac{\mu}{M_{\rm pl}} \right)^2 x_\ast^{2-p}.
\end{equation}
When $x_\ast \ll 1$ and $p<1$, we have $\epsilon \ll |\eta|$\footnote{
With the value of the model parameters assumed here, we have very
small $\epsilon$ as $\epsilon = 10^{-6} - 10^{-5}$.
}, which, with the help of Eq.~\eqref{eq:slow-roll} gives
the spectral index $n_s$ as 
\begin{equation}
%\label{ }
n_s \simeq 1 + 2 \eta \simeq 1 - \frac{2(1-p)}{(2-p)N}.
\end{equation}
If we take $p=0.8$, the spectral index is $n_s \simeq 0.993$  for
$N=50$,  which is outside the region allowed by Planck. However, when the curvaton generates a secondary inflation to
make $N$ much smaller, say $N=10$, the spectral index becomes $n_s
\simeq 0.966$, which gives a good fit to the current Planck data.

\begin{figure}[htbp]
\begin{center}
  \resizebox{160mm}{!}{
      \includegraphics{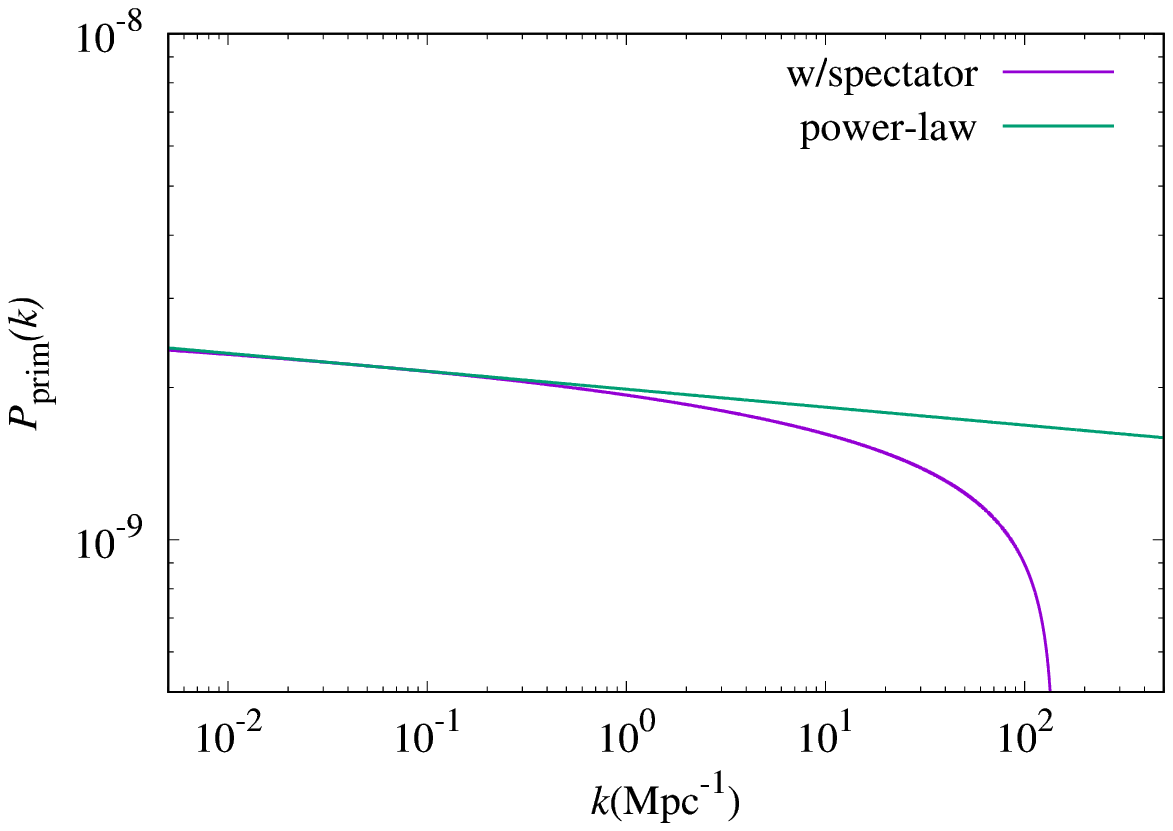} \hspace{5mm}
          \includegraphics{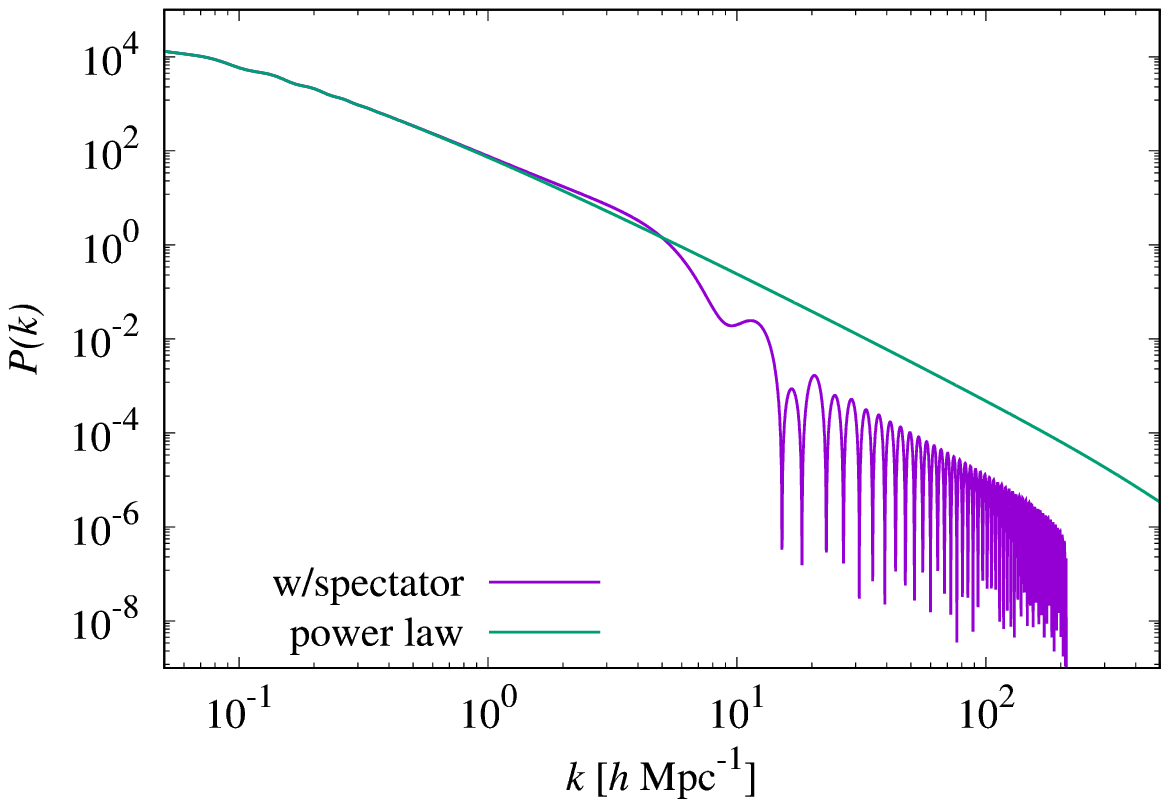}
  }
\end{center}
\caption{Primordial power spectrum (left) and linear matter power
  spectrum (right) for a spectator model with VHI model whose
  potential is given by Eq.~\eqref{eq:V_VHI}. Here we assume $p =
  0.84$. For the primordial power spectrum, there are two
  characteristic damping scales: $k_2$ and $k_{\rm end}$, which
  correspond to the modes that crossed the horizon at the time of the
  start of the second inflation and the end of the first inflation,
  respectively. In the matter power spectrum, the additional effect of
  oscillatory damping of the transfer function is visible. Depending
  on model parameters, the matter power spectrum can be either
  enhanced or suppressed at intermediate scales.}
\label{fig:power_spectrum}
\end{figure}

In Fig.~\ref{fig:power_spectrum}, we show the primordial power
spectrum (left) and matter power spectrum (right) for the case of
$p=0.84$ in the VHI inflation model, in which the CMB scale
corresponds to the modes exited the horizon at $N\sim 8$ during the
first inflationary period.  For comparison, the power spectrum is also
shown for the case where the power-law form $P_{\zeta} \propto k^{n_s
  - 1}$ is assumed.  As can be understood from
Eq.~\eqref{eq:zeta_phi1} in the previous section, the curvature
perturbations are generally suppressed on small scales where the modes
exit the horizon close to the end of inflation, since $V_\phi$ is
getting larger.  As discussed in the previous section, in our
scenario, small scale fluctuations of the modes which exited the
horizon during the second inflationary period are much smaller that
those on larger scales generated during the first inflation. Therefore
we can see the cutoff of the primordial power spectrum at around
$k_{\rm end} \sim 10^2 ~{\rm Mpc}^{-1}$ which corresponds to the mode
which exited the horizon at the end of the first inflationary
period. This can be clearly seen in the left panel of
Fig.~\ref{fig:power_spectrum}, which gives interesting implications to
the tension between CMB and subgalactic scales.

\begin{figure}[htbp]
\begin{center}
\vspace{10mm}
  \resizebox{100mm}{!}{
      \includegraphics{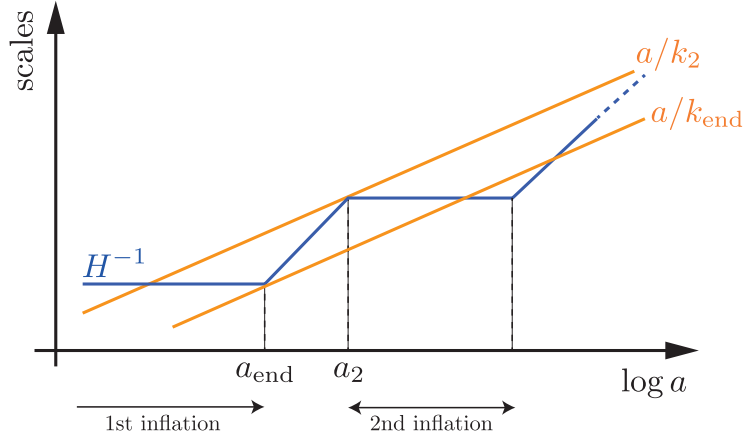}
  }
  
\end{center}
\caption{Schematic figure of corresponding scales during the two
  periods of inflation, the final radiation dominated era, and the
  (possible) intermediate radiation (or matter-like $\phi$
    oscillation) dominated phase.}
\label{fig:scale}
\end{figure}

To discuss the implication of our scenario for the small scale
structure, we calculate the matter power spectrum in the present
Universe.  In the right panel of Fig~\ref{fig:power_spectrum}, we plot
the matter power spectrum at $z=0$ for the same  model.  To
calculate the matter power spectrum, we need to incorporate the
effects of the evolution of fluctuations after the modes crossed the
horizon. In particular, there are two periods of inflation in our
model, which is quite similar to the case of thermal inflation where a
mini-inflation occurs after the first inflation driven by the
inflaton.  The transfer function in thermal inflation model has been
investigated in \cite{Hong:2015oqa}, where an analytic formula is
provided as
\begin{align}
\label{eq:transfer}
T (k) =& 
\cos \left[ \left( \frac{k}{k_2} \right) \int_0^{\infty} \frac{d\alpha}{\sqrt{\alpha ( 2 + \alpha^3)}}  \right] \notag \\
&
\qquad\qquad
+ 6 \left( \frac{k}{k_2} \right) \int_0^{\infty} \int^\gamma_0  d\beta \left( \frac{\beta}{2 + \beta^3} \right)^{3/2} 
\sin \left[ \left( \frac{k}{k_2} \right)  \int_\gamma^{\infty} \frac{d\alpha}{\sqrt{\alpha ( 2 + \alpha^3)}}  \right] \,.
\end{align}
Here  $k_2$ is the wave number which corresponds to the mode which
``touched" the horizon at the start of the second inflation ($k_2$ is
denoted as $k_b$ in \cite{Hong:2015oqa}).  
The above transfer function
exhibits oscillatory damping at scales smaller than $k_2$ which can be
related to $k_{\rm end}$, the mode which exited the horizon at the end
of the first inflation once we fix the background evolution.  Since $k
= aH$ holds at the time of the horizon crossing, one has
\begin{equation}
%\label{ }
\frac{k_2}{k_{\rm end}} =  \frac{a_2}{a_{\rm end}} \frac{H_2}{H_{\rm end}},
\end{equation}
where a subscript ``2" denotes that the quantity is evaluated at the
beginning of the second inflation.  
In Fig.~\ref{fig:scale},  a schematic figure of the horizon crossings for two characteristic scales $k_2$ and $k_{\rm end}$ is shown.
Assuming that the Universe is
$\phi$ oscillation-dominated\footnote{
Depending on the decay rate of the inflaton, the Universe may have
become radiation-dominated before the second inflation
starts. However, we assume that $\phi$ oscillation-domination in the
following.
}, in which $H \propto a^{-3/2}$, we obtain 
\begin{equation}
%\label{ }
k_2 = k_{\rm end} \left( \frac{H_2}{H_{\rm end}} \right)^{1/3}.
\end{equation}

Assuming that the Hubble parameter during inflation does not
  change much and hence we can approximate the Hubble parameter at the
  end of the first inflation as $H_{\rm end}^2 \simeq H_\ast^2 \simeq
  1.6 \times 10^{-7} \epsilon M_{\rm pl}^2$ as mentioned just above
  Eq.~\eqref{eq:sigma_ast}. The Hubble rate at the beginning of the
  second inflation can be written as $H_2^2 \simeq m_\sigma^2
  \sigma_\ast^2 /(3 M_{\rm pl}^2)$, from which one has
\begin{equation}
%\label{ }
k_2 \sim 0.01 ~k_{\rm end} \left( \frac{m_\sigma}{10^6~{\rm GeV}} \right)^{1/3} \left( \frac{\sigma_\ast}{10 M_{\rm pl}} \right)^{1/3} \left( \frac{10^{-5}}{\epsilon} \right)^{1/6}.
\end{equation}
For the case of $p=0.84$ depicted in Fig.~\ref{fig:power_spectrum},
the damping scale corresponding to the end of the first inflation is
given by $k_{\rm end} \sim 10^{2}~{\rm Mpc}^{-1}$, and hence the
damping scale in the transfer function is estimated as $k_2 \sim {\cal
  O}(1)~{\rm Mpc}^{-1}$.  

In the large and small scale limits ($k \ll k_2$ and $k \gg k_2$,
respectively), Eq.~\eqref{eq:transfer} gives the transfer function
\cite{Hong:2015oqa}
\begin{equation}
%\label{ }
T(k) \rightarrow
\begin{cases}
1  & (k \ll k_2), \\   \\
-\displaystyle\frac15 \cos \left( \nu_1 \frac{k}{k_2} \right) & (k \gg k_2),
\end{cases}
\end{equation}
where $ \nu_1 \simeq 2.2258 $. Cosmological N-body simulations with
the above transfer function have been studied by \cite{Leo:2018kxp}
for the thermal inflation model. It should be noted that, in our
model, the primordial power spectrum also gives the suppression on
small scales due to the mechanism discussed in the previous section,
and hence the power spectrum after the second inflation is given by
\begin{equation}
%\label{ }
\left. {\cal P}(k) \right|_{\rm after~2nd~inf.} = T^2(k) {\cal P}_\zeta (k), 
\end{equation}
where $T(k)$ is given by Eq.~\eqref{eq:transfer} and ${\cal P}_\zeta
(k)$ can be calculated by Eq.~\eqref{eq:P_prim_inflaton} when only the
inflaton contributes to the primordial curvature perturbation.

We input $ \left. {\cal P}(k) \right|_{\rm after~2nd~inf.} $ to the
public code {\tt CAMB} \cite{Lewis:1999bs} to obtain the matter power
spectrum at late time, which is shown in the right panel of
Fig.~\ref{fig:power_spectrum}.  As discussed above, there are two
damping scales in the model, $k_2$ and $k_{\rm end}$, corresponding to
the modes which exited the horizon at the end of the first inflation,
and those which ``touched" the horizon at the start of the second
inflation. This may give interesting consequences for structure
formation on subgalactic scales.

%%%%%%%%%%%%%%%%%%%%%%%%%%%%%%%%%%%%%%%%%%%%%%%%%%
\section{The power spectrum at small scales}\label{sec:small}

\begin{figure}[htbp]
\begin{center}
  \resizebox{140mm}{!}{
      \includegraphics{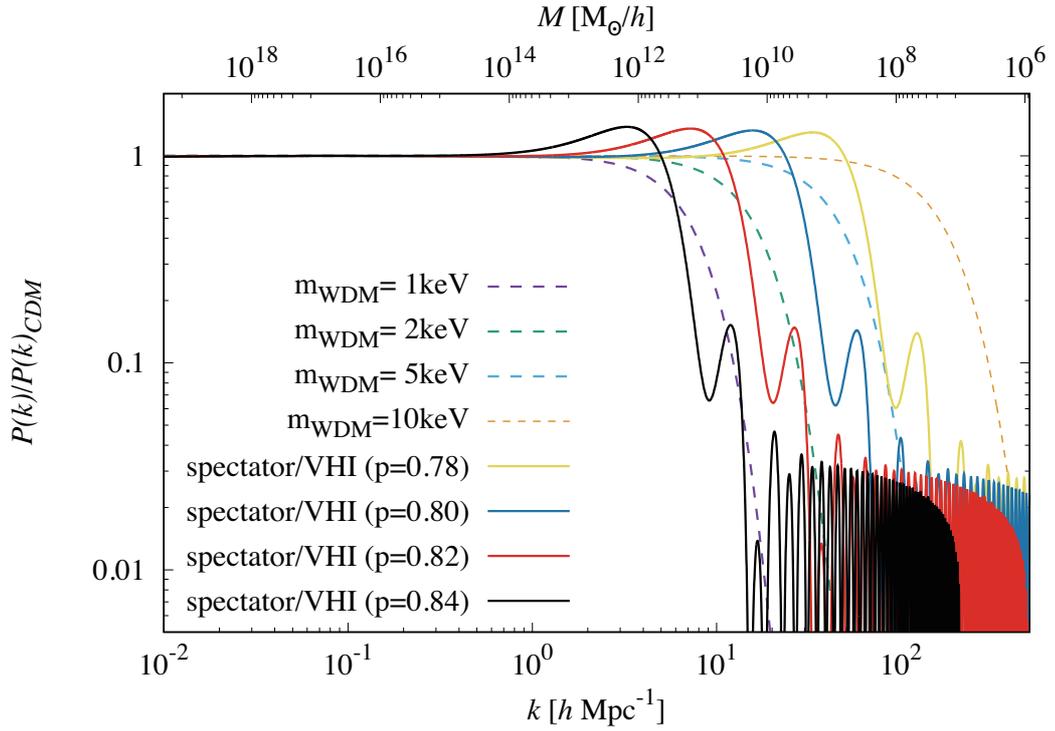}
  }
  
\end{center}
\caption{Ratio of the linear matter power spectra at $z=0$ between
  either spectator VHI models (solid lines) or warm dark matter models
  (dashed lines) relative to to the plain $\Lambda$CDM model. The top
  $x$-axis shows the equivalent mass, $M$ of a sharp $k$-space filter,
  indicative of the halo mass corresponding to a given length scale
  $k$.}
\label{fig:wdm_comparison}
\end{figure}

Alternatives to, or extensions of, the plain $\Lambda$CDM model are
most easily distinguished at small scales. Whereas plain $\Lambda$CDM
predicts bottom-up hierarchical structure formation after decoupling
on all observable scales, alternatives deviate in several
characteristic ways. For example, the free-streaming motions of
relativistic ``warm'' dark matter particles, such as sterile
neutrinos, dampen and erase initial perturbations, greatly reducing
the number of structures below a certain ``free streaming mass''
\cite{Bode:2000gq}. They also require that lower-mass halos and the
galaxies within them only form ``top down'' from the fragmentation of
more massive objects. Models where dark matter is coupled to photons
in the early Universe erase structures similar to warm dark matter,
but with some resonances at particular scales
(e.g. \cite{Boehm:2001hm, Schewtschenko:2015rno}). Models where dark
matter is self-interacting can change the inner structure of dark
matter halos, as well as lower the abundance of satellites
\cite{Vogelsberger:2012ku, Vogelsberger:2018bok}.

The most sensitive observations of small scale structures are the
abundance of present-day Local Group dwarf galaxies in halos of $\sim
10^{8}-10^{9.5}\Ms$, and structures observed in the Lyman-$\alpha$
forest at redshifts $z= 3-4$. Based on Lyman-$\alpha$ forest
measurements, warm dark matter models with thermal relics of less than
$3.5~{\rm keV}$ can be excluded at 99\% confidence level
\cite{Viel:2013apy}, but in the presence of Lepton asymmetry, the
correspondence between particle mass and structure damping differs
\cite{Shi:1998km}. However, the effects of WDM on structure formation
are similar in all models, and it is convenient to parameterise them
by the equivalent thermal relic mass.

For Local Group dwarf galaxies, several caveats exist: while a simple
comparison between the number of observed satellite galaxies, and the
number of halos appears to strongly disfavour plain $\Lambda$CDM,
astrophysical processes are understood to prevent the formation of
dwarf galaxies in low-mass halos, and also reduced the abundance of
low mass halos compared to collisionless simulations
(e.g. \cite{Sawala:2012cn}). Furthermore, the uncertain mass of the
Milky Way and of the Local Group has a significant effect on the
expected number of satellite halos. Indeed, for a massive Milky Way,
plain $\Lambda$CDM is strongly disfavoured, while for a low mass Milky
Way, many WDM models can be excluded \cite{Kennedy:2013uta}. In
simulations with Local Group analogues in the allowed mass range, it
appears that moderate WDM models are slightly favoured over CDM, but
observations of dwarf galaxies are not discriminant enough to
distinguish them \cite{Lovell:2016fec, Newton:2018izu}. A clearer
distinction may be possible based on the discovery or non-detection of
even lower mass dark matter halos, e.g. through their perturbations of
stellar streams, Milky Way halo stars, or via gravitational
lensing. While CDM predicts thousands of subhalos in the
$10^6-10^8\Ms$ range surrounding the MW, most WDM models predict very
few \cite{Sawala:2016tlo, Bose:2016irl}.

A modification of the initial power spectrum could act similar to warm
dark matter at the dwarf galaxy scale in the present Universe. In
Fig.~\ref{fig:wdm_comparison}, the ratio of the linear matter power
spectra at $z=0$ between the spectator VHI model and the plain
$\Lambda$CDM model, $P(k) / P(k)_{\rm CDM}$ is shown.  In both models,
the spectral index at the reference scale of $k=0.05~{\rm Mpc}^{-1}$
is taken as $n_s = 0.9645$ and other cosmological parameters are
assumed as $\Omega_b h^2 = 0.02225, \Omega_c h^2 = 0.1198, h=0.6727$
and $A_s =2.2 \times 10^{-9}$ \cite{Ade:2015xua}, where $\Omega_b$ and $\Omega_c$ are
density parameters for baryon and CDM, $h$ is the Hubble parameter and
$A_s$ is the amplitude of primordial power spectrum at the reference
scale. For the spectator VHI model, we have chosen the model
parameters, $\mu$ and $V_0$, for a fixed value of $p$ in such a way
that we obtain the above values of $n_s$ and $A_s$ at the reference
scale. By fixing the model parameters in this way, the number of
$e$-folds corresponding to the mode $k=0.05~{\rm Mpc}^{-1}$ are $N =
10.2, 9.4, 7.6$ and $7.8$ for $p=0.78, 0.8, 0.82$ and $0.84$,
respectively. For guidance, as dashed lines, we also include several
WDM models, assuming thermal relics of masses in the range
$1-10$~keV. It can be seen that, compared to WDM models, the power
spectrum in the spectator model begins to deviate at significantly
larger scales.

Depending on the model parameters, the reduction of the primordial
power, and the modification of the transfer function, discussed in
Section~\ref{sec:model}, can result in either a reduction or boost of
the linear matter power spectrum on galactic and supergalactic
scales. Below the cut-off scale $k_{\rm end}$, there is a sharp
suppression, more rapid than in the WDM case. Defining the ``filtering
mass'' as the scale where the abundance of halos is suppressed by half
relative to the standard model \cite{Bode:2000gq}, we see that the
spectator-VHI model with $p \sim 0.82$ has a similar filtering scale
to a WDM model with $m_{WDM} \sim 2$ keV, near the lower limits
allowed by Lyman-$\alpha$ observations \cite{Viel:2013apy}.

The scale-dependence of the WDM or VHI models relative to the standard
model, however, is quite different: We see that the WDM models change
the power by less than $1\%$ only half an order of magnitude above the
filtering scale, while in the spectator model, there is an even weaker
suppression on large scales, followed by a boost of $\sim 1/3$ on
intermediate scales. Models based on observed stellar kinematics,
abundance matching, and direct hydrodynamic cosmological simulations,
place the lowest-mass dwarf galaxies in dark matter halos around
$10^{8}-10^9\Ms$. While WDM solutions to the postulated plain
$\Lambda$CDM failures in this regime would affect only a very narrow
mass range, the signature of the spectator model would lead to a weak
change in galaxy abundance over a range of masses and affect different
scales in the Lyman-$\alpha$ forest. At the very low mass end, just
below the filtering scale, the spectator model predicts a very sharp
decline in power. For a filtering mass of $\sim 10^9\Ms$
(corresponding to the peak mass of typical Milky Way dwarf
spheroidals), the abundance of substructures below $10^7\Ms$ is
greatly reduced in the WDM model, but for the spectator model, such
substructures are non-existent. In this regime, where structures can
only be detected indirectly, even a very small number of detections
could therefore place significant constraints on spectator models.

In combination, this makes the spectator model clearly
falsifiable. Quantitative predictions will require full, cosmological
and hydrodynamic simulations, and will be the subject of future work.

%%%%%%%%%%%%%%%%%%%%%%%%%%%%%%%%%%%%%%%%%%%%%%%%%%
\section{Conclusion and Discussion} \label{sec:summary}

We have argued that, in models with a spectator field in the framework
of a small-field inflation, after the inflaton-driven
expansion there generically arises a second inflationary epoch which
is driven by the spectator field.  Depending on model parameters, the
second inflation can generate the large number of $e$-folds, possibly
even $N \sim 40-50$.  In this case, CMB scale fluctuations correspond
to the modes which exited the horizon at $N \sim 10$ when counted from
the end of the first inflation. Let us recall that this is a very
small number compared to the standard scenario where $N \sim 50 -60$.
Galactic scales correspond to the perturbations exiting around the
very end of the first inflation.

As we discussed in Section~\ref{sec:spectator}, in the case of
small-field inflation, the spectator fluctuations tend to give a
negligible contribution. As a result, primordial power spectrum on
small scales can be much suppressed compared to the standard plain
$\Lambda$CDM model while on large scales, the prediction can be
consistent with observations of CMB. This is  demonstrated in
Fig.~\ref{fig:wdm_comparison}.  Hence a model of the type with two
periods of inflation would give interesting implications for the
tension between small and large scale structure such as is manifest in
``missing satellites" and ``too big to fail" problems. As baryonic
physics strongly regulate galaxy formation on these scales, a
quantitative investigation of this issue will require cosmological
hydrodynamic simulations, which are left for future work.

Recent weak lensing results from KiDS \cite{Joudaki:2017zdt} continue
to yield $\sigma_8$ smaller than the value implied by the Planck data
by $2$ to $3 \sigma$. There are extensions of the plain $\Lambda$CDM
model devised to alleviate the apparent tension; these include
allowing for the curvature of the Universe, adopting dark energy
models with a time-varying equation of state, modifying general
relativity, assuming decaying dark matter (see e.g.,
\cite{Joudaki:2017zdt} for a recent discussion).

While the linear matter power spectrum shown in
Fig.~\ref{fig:wdm_comparison} also suggests a large-scale effect of
the modified transfer function, at low redshifts, this is likely to be
washed out due to mode coupling in the full, non-linear
evolution. More promising are observations of the 21cm forest around
the time of reionisation, which could be able to either detect or rule
out the characteristic bump in the power spectrum resulting from our
model. With SKA promising to make these measurements during the next
decade \cite{Koopmans:2015sua}, more detailed numerical studies of the
non-linear evolution in different inflation models seem to be
particularly timely.

%%%%%%%%%%%%%%%%%%%%%%%%%%%%%%%%%%%
\section*{Acknowledgments}
%%%%%%%%%%%%%%%%%%%%%%%%%%%%%%%%%%%
The authors also thank Baojiu Li, Matteo Leo, Mark Lovell and Kasper
Siilin for helpful comments and suggestions. T.T. would like to thank
the Helsinki Institute of Physics for the hospitality during the
visit, where this work was done.  T.S. is an Academy of Finland
Research Fellow and supported by grant number 314238. T.T is supported
by JSPS KAKENHI Grant Number 15K05084, 17H01131 and MEXT KAKENHI Grant
Number 15H05888.

%%%%%%%%%%%%%%%%%%%%%%%%%%%%%%%%%%%%%%%%%%%%%%%%%%

%%%%%%%%%%%%%%%%%%%%%%%%%%%%%%%%%%%%%%%%%%%%%%%%%%


\begin{thebibliography}{999}
%%%%%%%%%%%%%%%%%%%%%%%%%%%%%%%%%%%%%%%%%%%%%%%%%%


\bibitem{Klypin:1999uc}
  A.~A.~Klypin, A.~V.~Kravtsov, O.~Valenzuela and F.~Prada,
  %``Where are the missing Galactic satellites?,''
  Astrophys.\ J.\  {\bf 522} (1999) 82
%  doi:10.1086/307643
  [astro-ph/9901240].
 
  
\bibitem{Moore:1999nt}
  B.~Moore, S.~Ghigna, F.~Governato, G.~Lake, T.~R.~Quinn, J.~Stadel and P.~Tozzi,
  %``Dark matter substructure within galactic halos,''
  Astrophys.\ J.\  {\bf 524} (1999) L19
%  doi:10.1086/312287
  [astro-ph/9907411].

\bibitem{Larson:1974}
  Larson, R.~B.,
  % ``Effects of supernovae on the early evolution of galaxies,''
  Mon. Not. R. Astron. Soc. {\bf 169} (1974) 229.
%  doi:10.1093/mnras/169.2.229
  % adsurl = {http://adsabs.harvard.edu/abs/1974MNRAS.169..229L},
  
    \bibitem{Efstathiou:1992}
    Efstathiou, G.,
  % ``Suppressing the formation of dwarf galaxies via photoionization,''
    Mon. Not. R. Astron. Soc. {\bf 256} (1992) 43.
%    doi:10.1093/mnras/256.1.43P,
  % adsurl = {http://adsabs.harvard.edu/abs/1992MNRAS.256P..43E},



\bibitem{BoylanKolchin:2011de}
  M.~Boylan-Kolchin, J.~S.~Bullock and M.~Kaplinghat,
  %``Too big to fail? The puzzling darkness of massive Milky Way subhaloes,''
  Mon.\ Not.\ Roy.\ Astron.\ Soc.\  {\bf 415} (2011) L40
%  doi:10.1111/j.1745-3933.2011.01074.x
  [arXiv:1103.0007 [astro-ph.CO]].



\bibitem{Sawala:2015cdf}
  T.~Sawala {\it et al.},
  %``The APOSTLE simulations: solutions to the Local Group's cosmic puzzles,''
  Mon.\ Not.\ Roy.\ Astron.\ Soc.\  {\bf 457} (2016) no.2,  1931
%  doi:10.1093/mnras/stw145
  [arXiv:1511.01098 [astro-ph.GA]].


\bibitem{Gonzalez:2013pqa}
  R.~E.~Gonzalez, A.~V.~Kravtsov and N.~Y.~Gnedin,
  %``On the mass of the Local Group,''
  Astrophys.\ J.\  {\bf 793} (2014) 91
%  doi:10.1088/0004-637X/793/2/91
  [arXiv:1312.2587 [astro-ph.CO]].


\bibitem{Lovell:2016fec}
  M.~R.~Lovell {\it et al.},
  %``Properties of Local Group galaxies in hydrodynamical simulations of sterile neutrino dark matter cosmologies,''
  Mon.\ Not.\ Roy.\ Astron.\ Soc.\  {\bf 468} (2017) no.4,  4285
%  doi:10.1093/mnras/stx654
  [arXiv:1611.00010 [astro-ph.GA]].


\bibitem{Bozek:2018ekc}
  B.~Bozek {\it et al.},
  %``Warm FIRE: Simulating Galaxy Formation with Resonant Sterile Neutrino Dark Matter,''
  Mon.\ Not.\ Roy.\ Astron.\ Soc.\  {\bf 483} (2019) no.3,  4086
%  doi:10.1093/mnras/sty3300
  [arXiv:1803.05424 [astro-ph.GA]].


\bibitem{Schewtschenko:2015rno}
  J.~A.~Schewtschenko, C.~M.~Baugh, R.~J.~Wilkinson, C.~B\oe hm, S.~Pascoli and T.~Sawala,
  %``Dark matter\UTF{2013}radiation interactions: the structure of Milky Way satellite galaxies,''
  Mon.\ Not.\ Roy.\ Astron.\ Soc.\  {\bf 461} (2016) no.3,  2282
%  doi:10.1093/mnras/stw1078
  [arXiv:1512.06774 [astro-ph.CO]].



\bibitem{Viel:2013apy}
  M.~Viel, G.~D.~Becker, J.~S.~Bolton and M.~G.~Haehnelt,
  %``Warm dark matter as a solution to the small scale crisis: New constraints from high redshift Lyman-α forest data,''
  Phys.\ Rev.\ D {\bf 88} (2013) 043502
%  doi:10.1103/PhysRevD.88.043502
  [arXiv:1306.2314 [astro-ph.CO]].



\bibitem{Irsic:2017ixq}
  V.~Ir\v{s}i\v{c} {\it et al.},
  %``New Constraints on the free-streaming of warm dark matter from intermediate and small scale Lyman-$\alpha$ forest data,''
  Phys.\ Rev.\ D {\bf 96} (2017) no.2,  023522
%  doi:10.1103/PhysRevD.96.023522
  [arXiv:1702.01764 [astro-ph.CO]].




\bibitem{Kennedy:2013uta}
  R.~Kennedy, C.~Frenk, S.~Cole and A.~Benson,
  %``Constraining the warm dark matter particle mass with Milky Way satellites,''
  Mon.\ Not.\ Roy.\ Astron.\ Soc.\  {\bf 442} (2014) no.3,  2487
%  doi:10.1093/mnras/stu719
  [arXiv:1310.7739 [astro-ph.CO]].

\bibitem{Koopmans:2015sua}
  L.~V.~E.~Koopmans {\it et al.},
  %``The Cosmic Dawn and Epoch of Reionization with the Square Kilometre Array,''
  PoS AASKA {\bf 14} (2015) 001
  doi:10.22323/1.215.0001
  [arXiv:1505.07568 [astro-ph.CO]].

 
\bibitem{Bode:2000gq}
  P.~Bode, J.~P.~Ostriker and N.~Turok,
  %``Halo formation in warm dark matter models,''
  Astrophys.\ J.\  {\bf 556} (2001) 93
%  doi:10.1086/321541
  [astro-ph/0010389].

 
  \bibitem{Kamionkowski:1999vp} 
  M.~Kamionkowski and A.~R.~Liddle,
  %``The Dearth of halo dwarf galaxies: Is there power on short scales?,''
  Phys.\ Rev.\ Lett.\  {\bf 84}, 4525 (2000)
%  doi:10.1103/PhysRevLett.84.4525
  [astro-ph/9911103].
  
  \bibitem{Yokoyama:2000tz} 
  J.~Yokoyama,
  %``Inflation and the dwarf galaxy problem,''
  Phys.\ Rev.\ D {\bf 62}, 123509 (2000)
%  doi:10.1103/PhysRevD.62.123509
  [astro-ph/0009127].
  
\bibitem{Nakama:2017ohe} 
  T.~Nakama, J.~Chluba and M.~Kamionkowski,
  %``Shedding light on the small-scale crisis with CMB spectral distortions,''
  Phys.\ Rev.\ D {\bf 95}, no. 12, 121302 (2017)
%  doi:10.1103/PhysRevD.95.121302
  [arXiv:1703.10559 [astro-ph.CO]].
  

\bibitem{Enqvist:2014bua} 
  K.~Enqvist, T.~Meriniemi and S.~Nurmi,
  %``Higgs Dynamics during Inflation,''
  JCAP {\bf 1407}, 025 (2014)
%  doi:10.1088/1475-7516/2014/07/025
  [arXiv:1404.3699 [hep-ph]].

\bibitem{Enqvist:2014tta} 
  K.~Enqvist, S.~Nurmi and S.~Rusak,
  %``Non-Abelian dynamics in the resonant decay of the Higgs after inflation,''
  JCAP {\bf 1410}, no. 10, 064 (2014)
%  doi:10.1088/1475-7516/2014/10/064
  [arXiv:1404.3631 [astro-ph.CO]].
  
  \bibitem{Enqvist:2015sua} 
  K.~Enqvist, S.~Nurmi, S.~Rusak and D.~Weir,
  %``Lattice Calculation of the Decay of Primordial Higgs Condensate,''
  JCAP {\bf 1602}, no. 02, 057 (2016)
%  doi:10.1088/1475-7516/2016/02/057
  [arXiv:1506.06895 [astro-ph.CO]].


\bibitem{Enqvist:2001zp}
K.~Enqvist and M.~S.~Sloth,
%``Adiabatic CMB perturbations in pre big bang string cosmology,''
Nucl.\ Phys.\ B {\bf 626}, 395 (2002)
[arXiv:hep-ph/0109214].

\bibitem{Lyth:2001nq}
D.~H.~Lyth and D.~Wands,
%``Generating the curvature perturbation without an inflaton,''
Phys.\ Lett.\ B {\bf 524}, 5 (2002)
[arXiv:hep-ph/0110002].

\bibitem{Moroi:2001ct}
T.~Moroi and T.~Takahashi,
%``Effects of cosmological moduli fields on cosmic microwave background,''
Phys.\ Lett.\ B {\bf 522}, 215 (2001)
[Erratum-ibid.\ B {\bf 539}, 303 (2002)]
[arXiv:hep-ph/0110096].



\bibitem{Dvali:2003em}
  G.~Dvali, A.~Gruzinov, M.~Zaldarriaga,
  %``A new mechanism for generating density perturbations from inflation,''
  Phys.\ Rev.\  {\bf D69}, 023505 (2004).
  [astro-ph/0303591].


\bibitem{Kofman:2003nx}
  L.~Kofman,
  %``Probing string theory with modulated cosmological fluctuations,''
    [astro-ph/0303614].


\bibitem{Starobinsky:1986fx} 
  A.~A.~Starobinsky,
  %``Stochastic De Sitter (inflationary) Stage In The Early Universe,''
  Lect.\ Notes Phys.\  {\bf 246}, 107 (1986).
%  doi:10.1007/3-540-16452-9_6
  


\bibitem{Enqvist:2012xn}
  K.~Enqvist, R.~N.~Lerner, O.~Taanila and A.~Tranberg,
  %``Spectator field dynamics in de Sitter and curvaton initial conditions,''
  JCAP {\bf 1210}, 052 (2012)
%  doi:10.1088/1475-7516/2012/10/052
  [arXiv:1205.5446 [astro-ph.CO]].

\bibitem{Hardwick:2017fjo} 
  R.~J.~Hardwick, V.~Vennin, C.~T.~Byrnes, J.~Torrado and D.~Wands,
  %``The stochastic spectator,''
  JCAP {\bf 1710}, 018 (2017)
%  doi:10.1088/1475-7516/2017/10/018
  [arXiv:1701.06473 [astro-ph.CO]].




\bibitem{Langlois:2004nn}
  D.~Langlois and F.~Vernizzi,
  %``Mixed inflaton and curvaton perturbations,''
  Phys.\ Rev.\  D {\bf 70}, 063522 (2004)
  [arXiv:astro-ph/0403258].

\bibitem{Moroi:2005kz}
  T.~Moroi, T.~Takahashi and Y.~Toyoda,
  %``Relaxing constraints on inflation models with curvaton,''
  Phys.\ Rev.\  D {\bf 72}, 023502 (2005)
  [arXiv:hep-ph/0501007].

\bibitem{Ichikawa:2008iq}
  K.~Ichikawa, T.~Suyama, T.~Takahashi and M.~Yamaguchi,
  %``Non-Gaussianity, Spectral Index and Tensor Modes in Mixed Inflaton and
  %Curvaton Models,''
  Phys.\ Rev.\  D {\bf 78}, 023513 (2008)
  [arXiv:0802.4138 [astro-ph]].


\bibitem{Dimopoulos:2011gb}
  K.~Dimopoulos, K.~Kohri, D.~H.~Lyth and T.~Matsuda,
  %``The inflating curvaton,''
  JCAP {\bf 1203}, 022 (2012)  [arXiv:1110.2951 [astro-ph.CO]].




\bibitem{Lazarides:2004we} 
  G.~Lazarides, R.~R.~de Austri and R.~Trotta,
  %``Constraints on a mixed inflaton and curvaton scenario for the generation of the curvature perturbation,''
  Phys.\ Rev.\ D {\bf 70}, 123527 (2004)
  [hep-ph/0409335].



\bibitem{Moroi:2005np}
  T.~Moroi and T.~Takahashi,
  %``Implications of the curvaton on inflationary cosmology,''
  Phys.\ Rev.\  D {\bf 72}, 023505 (2005)
  [arXiv:astro-ph/0505339];



\bibitem{Fonseca:2012cj} 
  J.~Fonseca and D.~Wands,
  %``Primordial non-Gaussianity from mixed inflaton-curvaton perturbations,''
  JCAP {\bf 1206}, 028 (2012)
  [arXiv:1204.3443 [astro-ph.CO]].

  
\bibitem{Enqvist:2013paa}
  K.~Enqvist and T.~Takahashi,
  %``Mixed Inflaton and Spectator Field Models after Planck,''
  JCAP {\bf 1310} (2013) 034
  [arXiv:1306.5958 [astro-ph.CO]].


\bibitem{Vennin:2015vfa} 
  V.~Vennin, K.~Koyama and D.~Wands,
  %``Encyclop\UTF{00E6}dia curvatonis,''
  JCAP {\bf 1511}, 008 (2015)
%  doi:10.1088/1475-7516/2015/11/008
  [arXiv:1507.07575 [astro-ph.CO]].
    
\bibitem{Fujita:2014iaa} 
  T.~Fujita, M.~Kawasaki and S.~Yokoyama,
  %``Curvaton in large field inflation,''
  JCAP {\bf 1409}, 015 (2014)
  %doi:10.1088/1475-7516/2014/09/015
  [arXiv:1404.0951 [astro-ph.CO]].

\bibitem{Haba:2017fbi} 
  N.~Haba, T.~Takahashi and T.~Yamada,
  %``Sneutrinos as Mixed Inflaton and Curvaton,''
  JCAP {\bf 1806}, no. 06, 011 (2018)
%  doi:10.1088/1475-7516/2018/06/011
  [arXiv:1712.03684 [hep-ph]].


\bibitem{Ichikawa:2008ne} 
  K.~Ichikawa, T.~Suyama, T.~Takahashi and M.~Yamaguchi,
  %``Primordial Curvature Fluctuation and Its Non-Gaussianity in Models with Modulated Reheating,''
  Phys.\ Rev.\ D {\bf 78}, 063545 (2008)
%  doi:10.1103/PhysRevD.78.063545
  [arXiv:0807.3988 [astro-ph]].
  
  

\bibitem{Starobinsky:1986fxa}
  A.~A.~Starobinsky,
  %``Multicomponent de Sitter (Inflationary) Stages and the Generation of
  %Perturbations,''
  JETP Lett.\  {\bf 42} (1985) 152
  [Pisma Zh.\ Eksp.\ Teor.\ Fiz.\  {\bf 42} (1985) 124].


\bibitem{Sasaki:1995aw}
  M.~Sasaki and E.~D.~Stewart,
  %``A General Analytic Formula For The Spectral Index Of The Density
  %Perturbations Produced During Inflation,''
  Prog.\ Theor.\ Phys.\  {\bf 95}, 71 (1996)
  [arXiv:astro-ph/9507001]. 


\bibitem{Sasaki:1998ug}
  M.~Sasaki and T.~Tanaka,
  %``Super-horizon scale dynamics of multi-scalar inflation,''
  Prog.\ Theor.\ Phys.\  {\bf 99}, 763 (1998)
  [arXiv:gr-qc/9801017].
  
\bibitem{Martin:2013tda} 
  J.~Martin, C.~Ringeval and V.~Vennin,
  %``Encyclop\UTF{00E6}dia Inflationaris,''
  Phys.\ Dark Univ.\  {\bf 5-6}, 75 (2014)
%  doi:10.1016/j.dark.2014.01.003
  [arXiv:1303.3787 [astro-ph.CO]].

  \bibitem{Hong:2015oqa} 
  S.~E.~Hong, H.~J.~Lee, Y.~J.~Lee, E.~D.~Stewart and H.~Zoe,
  %``Effects of thermal inflation on small scale density perturbations,''
  JCAP {\bf 1506}, 002 (2015)
%  doi:10.1088/1475-7516/2015/06/002
  [arXiv:1503.08938 [astro-ph.CO]].


\bibitem{Leo:2018kxp} 
  M.~Leo, C.~M.~Baugh, B.~Li and S.~Pascoli,
  %``N-body simulations of structure formation in thermal inflation cosmologies,''
  JCAP {\bf 1812}, no. 12, 010 (2018)
%  doi:10.1088/1475-7516/2018/12/010
  [arXiv:1807.04980 [astro-ph.CO]].


\bibitem{Lewis:1999bs} 
  A.~Lewis, A.~Challinor and A.~Lasenby,
  %``Efficient computation of CMB anisotropies in closed FRW models,''
  Astrophys.\ J.\  {\bf 538}, 473 (2000)
%  doi:10.1086/309179
  [astro-ph/9911177].
 

\bibitem{Boehm:2001hm}
  C.~Boehm, A.~Riazuelo, S.~H.~Hansen and R.~Schaeffer,
  %``Interacting dark matter disguised as warm dark matter,''
  Phys.\ Rev.\ D {\bf 66} (2002) 083505
%  doi:10.1103/PhysRevD.66.083505
  [astro-ph/0112522].




\bibitem{Vogelsberger:2012ku}
  M.~Vogelsberger, J.~Zavala and A.~Loeb,
  %``Subhaloes in Self-Interacting Galactic Dark Matter Haloes,''
  Mon.\ Not.\ Roy.\ Astron.\ Soc.\  {\bf 423} (2012) 3740
%  doi:10.1111/j.1365-2966.2012.21182.x
  [arXiv:1201.5892 [astro-ph.CO]].


\bibitem{Vogelsberger:2018bok}
  M.~Vogelsberger, J.~Zavala, K.~Schutz and T.~R.~Slatyer,
  %``Evaporating the Milky Way halo and its satellites with inelastic self-interacting dark matter,''
%  doi:10.1093/mnras/stz340
  arXiv:1805.03203 [astro-ph.GA].



\bibitem{Shi:1998km}
  X.~D.~Shi and G.~M.~Fuller,
  %``A New dark matter candidate: Nonthermal sterile neutrinos,''
  Phys.\ Rev.\ Lett.\  {\bf 82} (1999) 2832
%  doi:10.1103/PhysRevLett.82.2832
  [astro-ph/9810076].



\bibitem{Sawala:2012cn}
  T.~Sawala, C.~S.~Frenk, R.~A.~Crain, A.~Jenkins, J.~Schaye, T.~Theuns and J.~Zavala,
  %``The abundance of (not just) dark matter haloes,''
  Mon.\ Not.\ Roy.\ Astron.\ Soc.\  {\bf 431} (2013) 1366
%  doi:10.1093/mnras/stt259
  [arXiv:1206.6495 [astro-ph.CO]].


\bibitem{Newton:2018izu} 
  O.~Newton, M.~Cautun, A.~Jenkins, C.~S.~Frenk and J.~C.~Helly,
  %``The Milky Way's total satellite population and constraining the mass of the warm dark matter particle,''
  arXiv:1809.09625 [astro-ph.GA].
  %%CITATION = ARXIV:1809.09625;%%



\bibitem{Sawala:2016tlo}
  T.~Sawala, P.~Pihajoki, P.~H.~Johansson, C.~S.~Frenk, J.~F.~Navarro, K.~A.~Oman and S.~D.~M.~White,
  %``Shaken and Stirred: The Milky Way's Dark Substructures,''
  Mon.\ Not.\ Roy.\ Astron.\ Soc.\  {\bf 467} (2017) no.4,  4383
%  doi:10.1093/mnras/stx360
  [arXiv:1609.01718 [astro-ph.GA]].



\bibitem{Bose:2016irl} 
  S.~Bose {\it et al.},
  %``Substructure and galaxy formation in the Copernicus Complexio warm dark matter simulations,''
  Mon.\ Not.\ Roy.\ Astron.\ Soc.\  {\bf 464}, no. 4, 4520 (2017)
%  doi:10.1093/mnras/stw2686
  [arXiv:1604.07409 [astro-ph.CO]].





\bibitem{Ade:2015xua}
  P.~A.~R.~Ade {\it et al.} [Planck Collaboration],
  %``Planck 2015 results. XIII. Cosmological parameters,''
  Astron.\ Astrophys.\  {\bf 594}, A13 (2016)
%  doi:10.1051/0004-6361/201525830
  [arXiv:1502.01589 [astro-ph.CO]].

 
 \bibitem{Joudaki:2017zdt} 
  S.~Joudaki {\it et al.},
  %``KiDS-450 + 2dFLenS: Cosmological parameter constraints from weak gravitational lensing tomography and overlapping redshift-space galaxy clustering,''
  Mon.\ Not.\ Roy.\ Astron.\ Soc.\  {\bf 474}, no. 4, 4894 (2018)
  doi:10.1093/mnras/stx2820
  [arXiv:1707.06627 [astro-ph.CO]].
  
 
 
  
  
    
  
  
%%%%%%%%%%%%%%%%%%%%%%%%%%%%%%%%%%%%%%%%%%%%%%%%%%
\end{thebibliography}
\end{document}